\documentclass[a4paper,11pt]{article}

\usepackage{graphicx}
\usepackage{url}
\usepackage{amsmath}
\usepackage{amssymb}
\usepackage{cases}
\usepackage{fancybox}
\usepackage{qtree}

\newtheorem{proposition}{\hspace{0.0in}{\bf Proposition}}
\def\done{\hspace*{\fill} \rule{1.8mm}{2.5mm} \\ }

\makeatletter
\def\bbordermatrix#1{\begingroup \m@th
  \@tempdima 4.75\p@
  \setbox\z@\vbox{%
    \def\cr{\crcr\noalign{\kern2\p@\global\let\cr\endline}}%
    \ialign{$##$\hfil\kern2\p@\kern\@tempdima&\thinspace\hfil$##$\hfil
      &&\quad\hfil$##$\hfil\crcr
      \omit\strut\hfil\crcr\noalign{\kern-\baselineskip}%
      #1\crcr\omit\strut\cr}}%
  \setbox\tw@\vbox{\unvcopy\z@\global\setbox\@ne\lastbox}%
  \setbox\tw@\hbox{\unhbox\@ne\unskip\global\setbox\@ne\lastbox}%
  \setbox\tw@\hbox{$\kern\wd\@ne\kern-\@tempdima\left[\kern-\wd\@ne
    \global\setbox\@ne\vbox{\box\@ne\kern2\p@}%
    \vcenter{\kern-\ht\@ne\unvbox\z@\kern-\baselineskip}\,\right]$}%
  \null\;\vbox{\kern\ht\@ne\box\tw@}\endgroup}
\makeatother

\begin{document}

\title{Fake View Analytics in Online Video Services}
\author{Liang Chen, Yipeng Zhou and Dah Ming Chiu}
\maketitle

\begin{abstract}
Online video-on-demand(VoD) services invariably maintain a view count for each video they serve, and it has become an important currency for various stakeholders, from viewers, to content owners, advertizers, and the online service providers themselves. There is often significant financial incentive to use a robot (or a botnet) to artificially create \emph{fake views}. How can we detect the fake views? Can we detect them (and stop them) using online algorithms as they occur? What is the extent of fake views with current VoD service providers? These are the questions we study in the paper. We develop some algorithms and show that they are quite effective for this problem.


\end{abstract}

\section{Introduction}
In recent years, Internet video streaming (or VoD) service has become very popular. The content includes movies (traditionally distributed via VCR rentals), news, sports and TV series (traditionally via TV broadcasting), and user generated content (UGC), like videos on YouTube. Invariably, these online video services maintain a \emph{view count} for each video. The view count is useful to different parties in the video streaming service. To viewers, it has a recommendation value. To content owners, the view count measures the popularity of a video relative to other videos, and helps to establish a value for the video. Popular videos are likely to draw more attentions. If a video is shown with advertisements, then view count may help determine service rendered for the advertizers.
In the online video service industry,  view count is considered as a currency. Therefore it is important to keep view count accurate and trust-worthy.

Is the video count always dependable? Apparently not. In December 2012, Google announced that they cut 2 billion fake views from some large record company sites, because some of their videos were found to be artificially inflated \cite{youtube}. As far as we know, this problem is quite prevalent among all VoD service providers. Since fake view count can directly translate to financial benefits, it attracts people to make special tools of different level of sophistication for view count manipulation. Such tools, or services to boost view counts, are sold openly on the market.

Video service providers are all aware of the problem with fake view counts. How they handle this problem is usually not public information. In some sense, inflated view count is not all bad for a video service provider, as additional view count inflates its own popularity. But there are good reasons for content providers to deal with this problem seriously. Truthful view counts are important for online video service operation. For example, view count helps service providers to determine the value, hence how much to pay for different videos. Fake views incur addition load for the provider which increases its operating cost. But far more importantly, the service provider needs to establish trust from its viewers and advertizers who help fund its operation.

How to judge whether a view is real or \emph{fake}, is actually not that straightforward. The duration of a view certainly is relevant, but you cannot say a view less than a certain duration is definitely fake. The number of short duration views within a short time also makes it more likely a fake view attack, but how many views within what duration? If a large number of short views are from the same user (IP address), it also makes it very likely they are fake, but again, what is the threshold for the number of views from the same user? What if the IP address is that of a NAT box, which aggregate traffic from many users? There are other illegitimate ways people use to attract views, for example, by using an inappropriate title. This makes the judgement of a fake view even more ad hoc. For Google's YouTube, it is well known that When the view count reaches 300, YouTube \emph{freezes} the view count and starts a validation process of the views count. It may take several hours to one day before the view count update resumes. For a large scale service such as YouTube, detecting fake views on a continuing basis can be very expensive, and Google's approach is trade-off between maintaining view count's accuracy and its cost. The exact method Google uses to validate the view count of a video, as far as we know, is proprietary.

In this paper, we report a comprehensive study of this problem. We are able to collaborate with one of the largest online video service providers - Tencent video \cite{qqvideo} - who is facing similar issue as YouTube and provides us access to the data needed for this study. We will first review the background, including a brief overview of the online VoD system and its scale, and a discussion of the whole ecosystem of the business of generating fake views - the kind of tools used, how the tools or services are provided, and the motivation for different parties. This will be done in Section 2, in which we will also give a more technical description and discussion of the techniques in generating fake views\footnote{Note, in this paper our definition of fake view is limited to views generated by some kind of robot rather than a normal view from a human viewer. In YouTube, they may consider all views of a video with a misleading title as fake views. While this is reasonable, the detection of this kind of fake views is very hard to do without human help.}. Our main results concern algorithms to detect fake views, which in turn can be considered effectively a definition of fake view. The results are reported in Section 3 and 4 for offline algorithms and online algorithms respectively. In Section 5, we give some statistics of the current extent of fake views in the system we studied. Finally, we give a discussion of related works in Section 6 and conclusion in Section 7.

\section{Background}



\subsection{The online video service under study}
\begin{figure}[ht]
\centering
\includegraphics[width=0.6\textwidth]{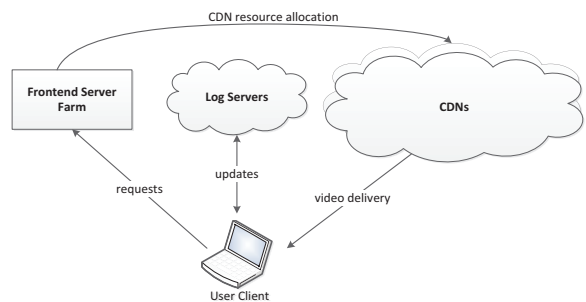}
\caption{General large-scale VoD service platform.}\label{fig:vod}
\end{figure}

Tencent Video is one of the largest video streaming service providers in China. They have more than 45 million active users on a daily basis, and more than 1.5 million users online concurrently during busy hours. Their video content includes movie, TV episodes, music/entertainment video, as well as many short clips of news, sports and user generated content (UGC). The online VoD service is delivered via HTTP/TCP, served by many servers in multiple CDN services. User viewing records are automatically collected using their infrastructure, and stored in a large data warehouse (consisting of log servers) as shown in Fig.~\ref{fig:vod}. Each view record is keyed by a video ID and IP address of the viewer. Our study is based on the analysis of these view records.

The view records are reported via HTTP messages by the Flash/HTML5 player embedded in the video webpages. This is a common way to collect information about user experience (QoE) for the online video service. Based on the reports, the VoD service provider calculates the view count for each video. The servers collecting the view records are well-connected with the access ISPs to ensure good coverage of all users. The collected data is quite massive, which makes our data analysis quite time and resource consuming.



\subsection{The ecosystem for fake views}
In the era of eyeball economy, attracting people's attention on Internet is a big business. One way to boost the view count (hence drawing attention) is to use some software tool to mimic the action of viewing a video. These tools are created by software hackers who learned enough about how the video system works to use robots (software program) to generate views. These tools are then sold on e-commerce sites such as Taobao (the equivalent of eBay in China). Depending on the sophistication of the tool, it may go for as little as 50 Yuan (or about \$ 8). Others use advanced tools to provide a service, which charges a certain amount of money for a given number of fake views.  For example, we notice one such service charges 10 Yuan (or \$ 1.6) per 10000 views for a video served by Tencent, YouKu or Tudou in China. Also, we find there are similar services for YouTube videos; for example, one service advertizes a charge of \$ 14 for creating 5000 YouTube views \cite{buyview}.

Obviously there is a market demand for these tools and services, from people who are hungry in boosting the popularity of their videos. The people who pay for fake views may not be (and usually are not) the pop stars featured in the videos, but their agents or promotors, or even fans. There are various other situations when people can profit from fake views, usually because the view count directly influences monetary settlement between different parties, for example in advertizement, or in joint content development contracts. For UGC, although the view count may not map to monetary settlement, some times there is also incentive to boost the view count, to simply generate attention. So there is a whole spectrum of different motives and requirements for generating fake views.

\subsection{Fake view attack methods}
We now try to explain how the fake view tools work. By understanding how fake views are generated, it helps us to figure out ways to deal with the problem. The methods we describe are \emph{off-the-shelf}. Anyone can find these by simply searching the Internet using keywords related to fake view. It is interesting to observe that such tools are sold openly, as a way for product promotion, sometimes with the claim that they do not disrupt the video service.


From what are advertized for sale on the market, we found two types of tools, with significant price differentiation. The simple tool makes a direct emulation of video viewing from a browser. The tool generates HTTP requests to actually download part of the video. This is repeated many times, to  generate large number of views. While this method makes a good emulation of the real viewing action, it takes up quite a bit of resources, and its speed of increasing view counts is limited. From our observation, this kind of tool typically generates on the order a few thousands of views per day.



The second method is more sophisticated. For the online video service in our study (perhaps others as well), the view count is incremented when the player sends a report about the viewing session to the log server, rather than when the user requests for the video initially. With this knowledge, the tool creator simply needs to figure out the format of the view report to send fake view reports repeatedly to generate fake views. Since this method avoids any video downloading, it can generate fake views much faster, at the rate of millions per day.


Fake view attacks can be launched using the above two methods from a single machine. Some times, more than one machines are used at the same time for the attacks. This may be realized by embedding the fake view tool in some malware distributed to user machines without them knowing about it (for example when users visit a web site implanted with such malware). Subsequently, the distributed malware programs can execute asynchronously (acting on their own), or act together under some central coordination, operating as what is commonly referred to a \emph{botnet}. In this case, if the first method is used, the malware tries to turn off the sound and place the video in the background to avoid the owner of the machine noticing its operation. When a large number of robots are involved, this becomes a DDOS attack. The detection and defense against DDOS is a challenging problem.

\begin{table}[ht]
\centering
\caption{Attack methods and their speed}\label{tab:attack}
\begin{tabular}{c|c|c}
\hline
 & Artificial Views & Forged Reports\tabularnewline
\hline
\hline
Single IP & $<$ 10k/day & $\sim$ 10m/day\tabularnewline
\hline
Multiple IPs & 100k $\sim$ 10m/day & $>$ 10m/day\tabularnewline
\hline
\end{tabular}
\end{table}

We summarize the methods in Table \ref{tab:attack}. Note, high speed of fake view generation is a double-edged sword - while more fake views can be generated quickly, it also makes detection of fake views easier. More sophisticated tools may try to randomize the timing of fake view requests and pace the requests at more reasonable rates to evade detection.




\subsection{Evidence for fake views}
\begin{figure}[th]
\centering
\includegraphics[width=0.6\textwidth]{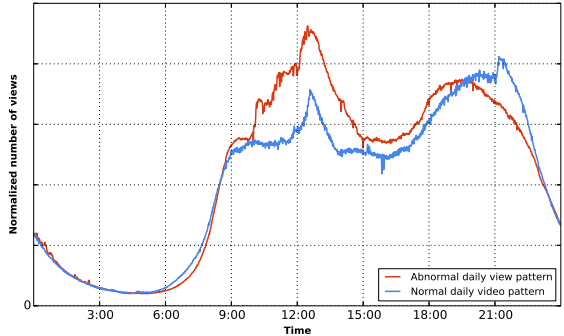}
\caption{Abnormal pattern of daily view count.}\label{fig:abnormal}
\end{figure}

An abnormal pattern in a daily view count plot gives us a first glance at the potential extent of the fake view problem. Normally, the daily view count arrival pattern is very similar, with the peak occurring some time during the busy hours, 18:00 $\sim$ 22:00 every day. We noticed one day that the peak arrival rate occurred at noon - see Fig.\ref{fig:abnormal}. By checking the view count reports around 12:00 on that day, we found that three videos collected around one million views in a short time. Those three videos effectively caused the peak arrival rate to shift to noon (instead of some time in the evening).

Based on further study of the log records, we found that the fake view attack on this day used forged view reports generated from multiple machines. Specifically, we found records of viewing reports sent to the log server from several different machines at random rates. The arrival of fake view reports was turned off and then back on irregularly, lasted around 3 hours. The aggregate rate of these fake view reports was around one million views within two hours on that day.

This incident led us to study more systematic methods for detecting fake views. Our main results concern offline detection, assuming all data in a day in the data warehouse are available for use. We will also briefly discuss online detection.


\section{Fake View Detection}
Generally speaking, there is no ground truth to definitively tell if a view request (or report) is real (initiated by a real user) or fake (initiated by robot). But many behavioral statistics can provide strong evidence to distinguish a fake view from a real one. So statistical analysis and machine learning methods are used to study this type of problems.


\subsection{Features useful for detection}
In machine learning, a \emph{feature} is an individually measurable heuristic property of a phenomenon being observed \cite{wiki:feature}. By choosing discriminating and independent features, different scenarios (e.g. real of fake view) can be \emph{classified} based on the features.

For detecting fake views, the most useful features include the number of times a particular video has been viewed by a user, within a short period of time. For example, if a user views a particular video for over a thousand times within an hour, it may quite reasonable to conclude that that user is in the process of generating fake views, most likely using a robot rather than manually, given the high speed.  The problem, however, is that the online video services typically do not require user authentication for accessing videos, at least of free services. Besides, there is always privacy concerns with tracking users on the Internet.

Without user identification, we can collect behavioral statistics of requests and reports sent in from the same IP address. The IP-address based features, however, are not as reliable as user-based features for fake view detection. One reason is that a single public IP address may be shared by many users because of the use of Network Address Translation (NAT). Furthermore, IP addresses are often dynamically allocated to users by ISP; so even during the course of a day, a user may have used multiple different public IP addresses to connect to the Internet. Much of our results in this paper are about how to reliably use IP address-based features to classify fake views.

\begin{figure}[ht]
\centering
\includegraphics[width=0.6\textwidth]{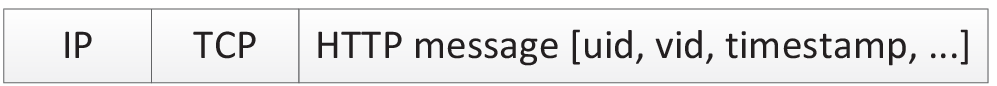}
\caption{View records sent via HTTP messages.}\label{fig:message}
\end{figure}

Since the view requests and view reports are sent as HTTP messages over TCP, the IP address in our records can be considered authentic. Because TCP requires setting up a connection, the IP address cannot be easily spoofed (UDP is vulnerable to IP spoofing, but it is rarely used to support the online video reporting). With the exception of the IP address, the content of the view report can be spoofed (rather than from a real viewing session generated by the player). This is the forged report method that we discussed earlier.

Besides the features associated with user identity and IP address, some attributes of a video can can be good features for detection use. For example, 1) the type of video: User Generated Content (UGC) videos are more prone to fake views than others because more content owners are likely to promote their videos; 2) the release time of the video: a newer video tends to have more reasons for promotion. Other features related to videos include votes, rating, and comments on videos.

\subsection{Analysis based on user's video access}


Let us first consider the problem assuming we know the video access pattern by users. A simple form of the data can be represented by the following matrix:
\[
A_{m \times n} = \bbordermatrix{~ & \text{vid}_1 & \text{vid}_2 & \ldots & \text{vid}_n \cr
                 \text{user}_1    &    a_{11}    &    a_{12}    & \ldots & a_{1n}       \cr
                 \text{user}_2    &    a_{21}    &    a_{22}    & \ldots & a_{2n}       \cr
                 \vdots           &    \vdots    &   \vdots     & \ddots & \vdots       \cr
                 \text{user}_m    &    a_{m1}    &    a_{m2}    & \ldots & a_{mn}
                 }
\]
Each row represents the video access pattern of a particular user, and each column represents the distribution of user views for a particular video. The element $a_{ij}$ is the number of views by the $i^{th}$ user of the $j^{th}$ video in a given day. Given this data, the most primitive method for detection is to set a threshold $K$ as too many views by a single user in a single day. In other words, $a_{ij} > K$ implies user $i$ is creating fake views for video $j$.

Although as we mentioned earlier that user access records are not generally available, there is a small fraction of users who happen to have logged into the system for some other reasons (e.g. social network services) before they access the online video service, and in this case, the video access records keep the user ID information. We can consider this fraction of the users a random sample of the whole user population, and see how well detection can work given user's video access information\footnote{It is possible that people engaged in fake view generation are more careful about logging into the system, but from the data, it does not seem that is the case.}.


For this small study, we derive the ground truth manually. For most of the users, it is either clear that they are normal users because the relative small number of viewing sessions they had during the day;
or that they are really robots because their video viewing history is humanly unlikely (such as viewing thoughts of videos in a day). For those users there is some ambiguity, we examined the viewing history manually to make a judgement whether the user is normal or abnormal. The manual review includes considering the rate at which the views are generated, the different number of videos viewed and other behavioral patterns.

\begin{figure}[ht]
\centering
\includegraphics[width=0.6\textwidth]{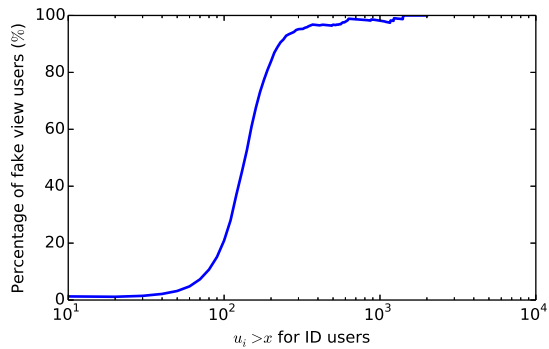}
\caption{Percentage of fake view users for ID users.}\label{fig:fakeuser}
\end{figure}

In Fig. \ref{fig:fakeuser}, we plot the percentage of fake view users (based on manual classification) against the number of daily views. We observe that fake view users account for 98.2\% or more for those who have larger than 1000 total views in a day; and the percentage is 95.2\% for viewers who have more than 300 daily views. In other words, a user with more than 300 daily views is very likely a fake view user.

\begin{figure}[ht]
\centering
\includegraphics[width=0.6\textwidth]{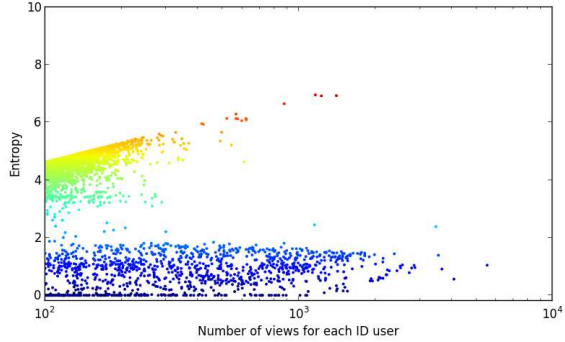}
\caption{Entropy for each user identified by ID.}\label{fig:eqq}
\end{figure}

The manual reviewing process inspired us to use the \emph{entropy} function to help classify users. The idea is quite simple. Normal users tend to view a few different videos during a day. Because of the variety of videos viewed, the entropy cannot be too low. On the other hand, if a user made a lot of views of the same video, the lack of variety means the entropy value is much lower. The exact equation is as shown below:
\[
H_u(i) = -\sum_{j=1}^{n}\frac{a_{ij}}{u_i} \ln \frac{a_{ij}}{u_i}\cdot I(a_{ij}>0),
\]
where $u_i=\sum_j a_{ij}$ and $I(a_{ij})$ is an indicator with value $1$ if $a_{ij}>0$ and $0$ otherwise.
The entropy function provides a good indication about a user's viewing behavior. We computed the entropy for every user and plot it in Fig.~\ref{fig:eqq}. Each user is represented by a dot. The x-axis value is the total number of views for the user and the y-axis is the entropy computed using the equation above. As evidenced from the plot, users can be roughly classified into two types. For the first type, their entropy values are above 2 and increase with the number of views, but the total user view count is mostly fewer than $300$; whereas the second type of users have low entropy values and many have large total user view counts.  By combining and analyzing the results in Fig.~\ref{fig:eqq} and Fig~\ref{fig:fakeuser}, we find that the first type users with large entropy are normal viewers, while the second type of users with smaller entropy are most often involved in generating fake views. This conclusion is quite intuitive. Those users generating fake views tend to access the same video (for fake views) all the time, hence their access pattern (distribution) is quite predictable, which is reflected in having low entropy. We can thus conclude that entropy is an effective technique for detecting fake views.

\begin{table}[ht]
\centering
\caption{Manually checking ID users with views $>$ 100}\label{tab:manual}
\begin{tabular}{c|c|c|c}
\hline
 & $H_{u}=0$ & $H_{u}<1$ & $H_{u}<3$\tabularnewline
\hline
\hline
\# of users & 166 & 649 & 1171\tabularnewline
\hline
percent of anomaly & 100\% & 100\% & 98.2\%\tabularnewline
\hline
\end{tabular}
\end{table}



\subsection{Analysis based on IP's video access}

Although applying entropy to user viewing patterns is a good technique for detecting fake views, we cannot rely on having user IDs all the time, for a variety of reasons discussed before.  For each viewing session report, the source IP address is always part of the record. The question is, can we substitute user ID with IP address?

The user ID does not map one-to-one with IP address. \emph{Network Address Translation} (NAT) boxes are widely used in the Internet for ease of management conserving public IP addresses. This is especially true in China. This means many users may share a single public IP address (of the NAT), as visible by the online video provider. Even when there is no NAT between the user and the video service, the user's IP address may be dynamically assigned. Therefore, we need to re-evaluate and think of new techniques for detecting fake views.





By slight abusing of notations, we can redefine the user access matrix to be the IP access matrix:

\[
A_{m \times n} = \bbordermatrix{~ & \text{vid}_1 & \text{vid}_2 & \ldots & \text{vid}_n \cr
                 \text{IP}_1    &    a_{11}    &    a_{12}    & \ldots & a_{1n}       \cr
                 \text{IP}_2    &    a_{21}    &    a_{22}    & \ldots & a_{2n}       \cr
                 \vdots           &    \vdots    &   \vdots     & \ddots & \vdots       \cr
                 \text{IP}_m    &    a_{m1}    &    a_{m2}    & \ldots & a_{mn}
                 }
\]
Similar as before, each row represents an IP's video requests, and each column represents the requests from different IPs for a video. The element $a_{ij}$ is the number of views from $i^{th}$ IP to $j^{th}$ video in a given day.
We define IP entropy based on the IP access matrix as follow:
\[
H_w(i) = -\sum_{j=1}^{n}\frac{a_{ij}}{w_i} \ln \frac{a_{ij}}{w_i}\cdot I(a_{ij}>0),
\]
where $w_i=\sum_j^n a_{ij}$ and $I(a_{ij})$ is an indicator with value $1$ if $a_{ij}>0$ and $0$ otherwise. Note, the number of IPs is much greater than the number of users with IDs known to us. For a particular day (9-4-2013), for example, the number of IPs ($m$) is $15,649,601$.
For a large fraction of IPs (69\% or $10,814,644$), $w_i \leq 3$ in a day, and their total views account for only around 21\% of all daily views. The IPs with few total views are unlikely to generate any fake views. For our studies, we decided to filter out all IPs with fewer than $50$ total views, leaving us with $m=140,341$, about 0.9\% of all IPs.

\begin{figure}[ht]
\centering
\includegraphics[width=0.6\textwidth]{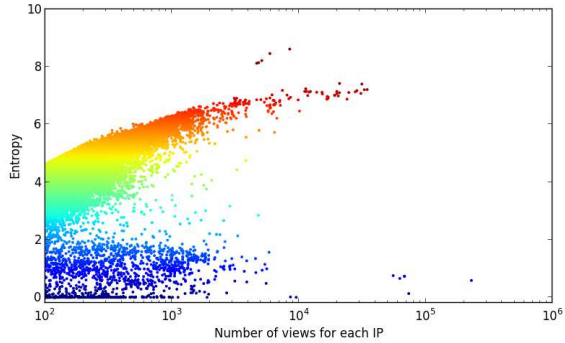}
\caption{Entropy for each IP with different number of views.}\label{fig:eip}
\end{figure}

The entropy of each IP is calculated and plotted in  Fig.~\ref{fig:eip}. Again, the entropy function helps us classify the IPs into two categories. Compared with Fig.~\ref{fig:eqq}, there is an important difference. Due to the existence of NAT, there are significant number of IPs with both large view counts and IP entropy.

Since there could be multiple users behind a single IP, we need to better understand the behavior of view count distribution generated by multiple users. For this purpose, we need to introduce \emph{video popularity}, denoted by $(\eta_1,\dots,\eta_n)$. $\eta_j$ is the probability for any view to be for video $j$. By definition, $\sum_{j=1}^n \eta_j = 1$. We now consider and compare IP entropy for two cases. In case one, users and IPs are one to one mapping, i.e. an IP is assigned to a unique user. In this case, the IP entropy should be similar to the user entropy. For users, it is rare to view the same video more than once. With such behavior, we can conclude that
\begin{proposition}
In our online VoD system, if there is a unique user behind each IP address, then the IP entropy $H_w(i)$ for a typical user $i$ increases logarithmically with the view count $w_i$, i.e.
\[ H_w(i) = \ln w_i.\]
\end{proposition}
It is not difficult to see this by assuming that the $w_i$ views are for a distinct set of $w_i$ videos. The user entropy in Fig.~\ref{fig:eqq} validates this property.

The challenge of anomaly detection is for case two, when users map many-to-one to IPs, which is the NAT case. In this case, what is the expected behavior of the IP entropy function?
Let us first consider the situation when a huge number of users are behind each IP address, so $w_i\gg n$. From law of large numbers, the distribution of $(a_{i1}/a_i, \dots, a_{in}/a_i)$ will approach $(\eta_1, \dots, \eta_n)$. This yields $E[H_w(i)] \geq -\sum_{j=1}^n\eta_j \ln \eta_j$. On the other hand, entropy is a convex function. So we would expect $H_w(i)$ to decrease and gradually approach a limit value determined by video popularity, $E[H_w(i)]$. However, from our measurement results Fig~\ref{fig:eip}, the IP entropy is increasing. This is because the number of viewed videos by a single IP is much less than the video population, i.e. $w_i\ll n$. In other words, there are a lot of unwatched videos from a single IP's point of view. As we increase the view count, it is highly likely that the additional view is for a perviously unwatched video. As we increase the number of watched video by an IP, we prove the following:
\begin{proposition}
As view count increases, whenever it increases the number of watched videos, the IP entropy is also increased.
\end{proposition}

\noindent{\bf Proof:}
To prove it, let us consider IP $i$. Assume that there is an unwatched video $k$ that will be requested by the next view. $\overrightarrow{a}_i$ represents the vector of current view counts, i.e. $\overrightarrow{a}_i =(a_{i1},\dots, a_{in})$. Let $e_{k}$ be a unit vector with value $1$ at the $k^{th}$ position and $0$ at all other positions. The IP entropy for IP $i$ with the incremental view is denoted by $H_w(\overrightarrow{a}_i+e_k)$ (again slightly abusing the notations), then
\begin{eqnarray*}
\!\!\!\!&    \!\!\!\! & H_w(\overrightarrow{a}_i+e_k)\\
 \!\!\!\!&  =  \!\!\!\! & -\sum_{j=1}^{n}\frac{a_{ij}}{w_i+1}\ln \frac{a_{ij}}{w_i+1}\cdot I(a_{ij}>0) - \frac{1}{w_i+1}\ln \frac{1}{w_i+1},\\
   \!\!\!\!& =  \!\!\!\!& -\frac{w_i}{w_i+1}\sum_{j=1}^{n}\frac{a_{ij}}{w_i}\big(\ln \frac{w_{i}}{w_i+1}+\ln \frac{a_{ij}}{w_i}\big)\cdot I(a_{ij}>0) \\
   \!\!\!\!&   \!\!\!\! &-\frac{1}{w_i+1}\ln \frac{1}{w_i+1},\nonumber\\
    \!\!\!\!& =  \!\!\!\!& \frac{w_i}{w_i+1} H_w(i) -\frac{w_i}{w_i+1}\ln \frac{w_i}{w_i+1}-\frac{1}{w_i+1}\ln \frac{1}{w_i+1}.
\end{eqnarray*}
From information theory, $H_w(i)\leq \ln w_i$, so
\begin{eqnarray*}
\!\!\!\!&    \!\!\!\! & H_w(\overrightarrow{a}_i+e_k)-H_w(\overrightarrow{a}_i)\\
 \!\!\!\!&  =  \!\!\!\! & -\frac{1}{w_i+1}H_P(\overrightarrow{a}_i)-\frac{w_i}{w_i+1}\ln \frac{w_i}{w_i+1}-\frac{1}{w_i+1}\ln \frac{1}{w_i+1}\\
 \!\!\!\!&  \geq  \!\!\!\! & \frac{1}{w_i+1}\ln w_i -\frac{w_i}{w_i+1}\ln \frac{w_i}{w_i+1}-\frac{1}{w_i+1}\ln \frac{1}{w_i+1} \\\!\!\!\!&  =  \!\!\!\! &\ln \frac{w_i+1}{w_i}>0
\end{eqnarray*}
Thus, the entropy is increasing.
\done

\begin{figure}[ht]
\centering
\includegraphics[width=0.6\textwidth]{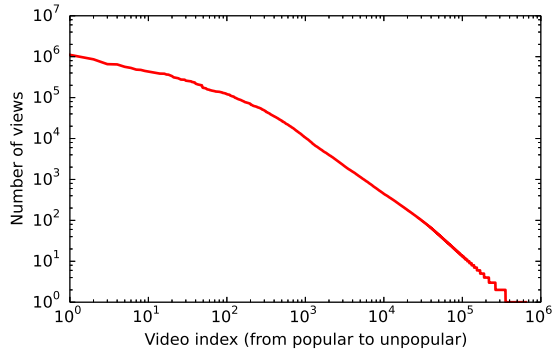}
\caption{Video popularity.}\label{fig:vid_pop}
\end{figure}

Therefore, we can conclude that the IP entropy increases with view count as long as $w_i\ll n$. However, it will eventually approach a limit. The speed depends on the skewness of the distribution of video popularity. As shown in Fig.~\ref{fig:vid_pop}, the video popularity is very skewed, which predicts that the IP entropy curve will not stay a straight line on the log plot, as is the case in Fig~\ref{fig:eip}.

Although we are able to characterize the normal IP entropy curve, this knowledge does not help detect IPs with fake views, since they do not necessarily have low IP entropy. For example, some user behind a large NAT generating $\delta$ fake views for some video $k$ may not decrease the IP entropy for that NAT's IP address sufficiently. The IP entropy depends on the distribution of $(\frac{a_{i1}}{w_i+\delta}, \dots,\frac{a_{ik}+\delta}{w_i+\delta}\dots)$. So, it is difficult to discover the anomalous IPs purely relying on IP entropy. Next we take a detour to consider how to detect videos with fake views first.

\subsubsection{Detecting videos with fake view}

In addition to the IP entropy, which tells us what different videos an IP accesses, we can define video entropy similarly, which tell us how different users request for a video. Video entropy can be used to classify videos with fake views, which can in turn help identify IPs with fake views as we show later.

Actually, tracking true video popularity is necessary for content providers as part of their business. For our collaborator for instance, the number of videos viewed in a day is routinely more than a million. In order to find out true video popularity, it is necessary to remove fake views.

Based on the IP access matrix, the video entropy is defined as
\[
H_v(j) = -\sum_{i=1}^{m}\frac{a_{ij}}{v_j} \ln \frac{a_{ij}}{v_j}\cdot I(a_{ij}>0),
\]
where $v_j=\sum_i^m a_{ij}$ is the total number views for video $j$. To study the video entropy, we introduce the concept of IP popularity. $\tau_i$ is the probability for any given view coming from IP $i$. Obviously, the value of $\tau_i$ depends on the number of users behind IP $i$ and $\sum_{i=1}^m \tau_i = 1$. Similar to the analysis of IP entropy, we can derive two conclusions for video entropy which are stated as propositions:
\begin{proposition}
If $v_j\gg m$, the video entropy for video $j$ is a decreasing function that approaches the limit $-\sum_{i=1}^{m}\tau_i\ln \tau_i$.
\end{proposition}
But similar to the IP entropy case, for most videos $v_j \ll m$, so we have:
\begin{proposition}
For a video $j$ and an incremental view from some IP $k$, if $k$ has not viewed $j$ before, then the video entropy's value is increased.
\end{proposition}
The proofs are very similar to that of the IP entropy case, so they are omitted. For the practical system we measured, indeed $v_j \ll m$, so on average the video entropy $H_v(j)$ increases with total number of views $v_j$, for a typical video $j$.

We analyzed some popular videos with a large number of daily requests. Such popular videos normally attract similarly large number of visitors from a diverse set of regions and IP addresses, making the video entropy high. If on the contrary, a popular video gets most of its view count from a few IPs, it is very likely that some or all of these views are fake views. Thus, entropy is again helpful to distinguish normal videos with fake view videos.
\[
\text{video}
\left\{
\begin{aligned}
&\text{fake views}\left\{\begin{aligned}
                     &\textit{single user}\\
                     &\textit{multiple users}
                     \end{aligned}\right.\\
&\text{normal views}
\end{aligned}\right.
\]

Thus for all videos, we can divide them into videos with (sufficient number of) fake views, or with normal views only. Out of videos with fake views, we can then detect if they are generated by a single IP or multiple IPs.

\begin{figure}[ht]
\centering
\includegraphics[width=0.6\textwidth]{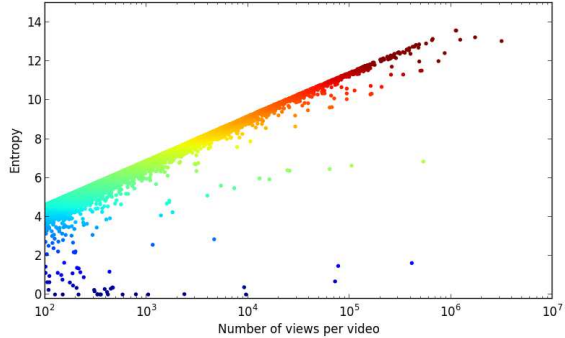}
\caption{Entropy for videos with different number of views.}\label{fig:evid}
\end{figure}

Based on the definition above, video entropy for a given day is calculated and plotted in Fig. \ref{fig:evid}.

\begin{figure}[ht]
\centering
\includegraphics[width=0.6\textwidth]{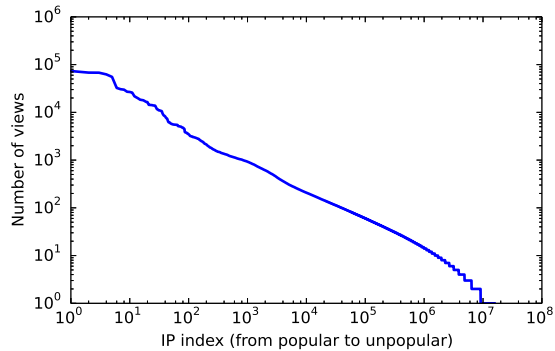}
\caption{IP popularity.}\label{fig:ip_pop}
\end{figure}

Perhaps not surprisingly, this figure shares many common features with the user entropy figure and IP entropy figure. Similarly, all videos can be roughly classified into two categories. The entropy of the first category of videos increases with the total number of views while the increase in entropy for the second type of videos is very slow or non-existent. Besides the similarity, we observe two differences. First, the video entropy for the first type increases more sharply than the IP entropy with number of views. Secondly, the fraction of the videos with low entropy is much smaller than the case for IPs. The first difference can be explained by the skewness of IP popularity distribution, which is plotted in Fig.~\ref{fig:ip_pop}. Compared with Fig.~\ref{fig:vid_pop}, clearly IP popularity is less skewed than video popularity. It is quite intuitive that this leads to the sharper increase of the video entropy.
The second difference can be explained by the fact that often multiple IPs are involved in generating fake views for a common video. This leads to more fake view IPs than fake view videos.

Based on the observation of Fig.~\ref{fig:evid} and the above analysis, we conclude that the a video $j$ with entropy approximately equal to $\ln v_j$ is a normal video without fake views; while videos with much lower entropy than the normal videos at about the same number of views are suspected to be fake view videos.




Videos with entropy close to upper limit or close to $0$ are easy for classification. The videos with entropy values in between are the difficult cases to handle automatically. We propose to use off-the-shelf machine learning techniques to further help us detect the fake view videos, especially for the difficult cases. Since the problem can be considered as a standard classification problem, the supervised learning approaches can be adopted. Supervised learning requires labeled training data to infer a function for mapping new examples. In our case, however, the daily viewed video is in the millions, which makes it hard to label each video (fake view video or not) manually for the training data preparation. Therefore, we have to resort to the semi-supervised learning approach to deal with our problem. Based on the features we have observed, the number of views and the entropy of video, we are able to detect the view count anomaly correspondingly.

\begin{figure}[ht]
\centering
\includegraphics[width=0.6\textwidth]{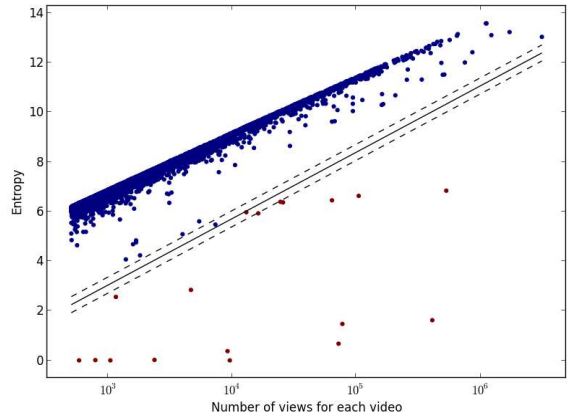}
\caption{TSVM classification for fake view video detection.}\label{fig:svm_vid}
\end{figure}

The semi-supervised learning makes use of both labeled and unlabeled data for training. This characteristic is quite suitable for our case: a small amount of labeled data with a large amount of unlabeled data. We manually classify 10 thousand videos by labeling them +1 and -1 as for the normal view and fake view videos respectively. To train the classifier, We use the transductive support vector machines (TSVM) \cite{vv98} as our detection model. Our TSVM algorithm is implemented based on the SVMLight toolbox \cite{jt99mit}. We derive the hyperplane from the model by processing our labeled and unlabeled training data. As shown in Fig. \ref{fig:svm_vid}, the hyperplane separates the normal view and fake view videos with linear kernel. We can also replace it with polynomial kernel to smooth the hyperplane. In practice, the linear one is efficient for anomaly detection.

\begin{figure}[ht]
\centering
\includegraphics[width=0.6\textwidth]{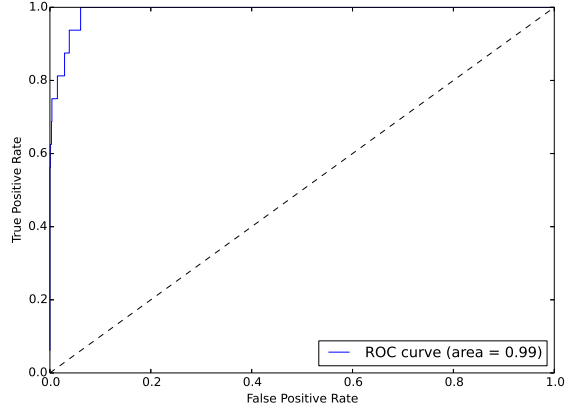}
\caption{ROC curve for fake view video detection.}\label{fig:roc_vid}
\end{figure}

With the test datasets, we examine the effectiveness of our TSVM classifier. We prepared one week's data, and use the model to detect fake view videos. By checking the false positive and false negative rates of the classification results (linear kernel), we find that the TSVM approach is quite good as shown in Fig~\ref{fig:roc_vid}. The area under ROC curve is around 0.99. The error usually occurs for videos located at the bottom-left corner which have small view counts.

\begin{figure}[ht]
\centering
\includegraphics[width=0.65\textwidth]{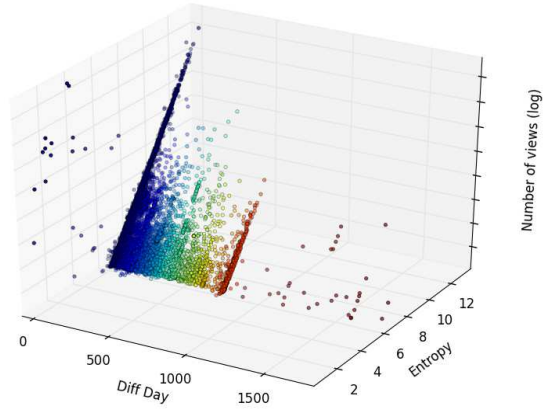}
\caption{Adding the day difference feature for fake view detection.}\label{fig:svm3}
\end{figure}

For daily fake view examination, we introduce one more feature, the release date of a video, to help further improve accuracy. We calculate the date difference between the day of daily view records and the day the video is released. We observe the three features in Fig. \ref{fig:svm3} with different colors depicting the date difference (darker ones are nearer to present time). It shows that the fake views are usually generated within a short time (around ten to thirty days) after the release date of a video.  It indicates the fake view creators have strong incentive to inflate the view count soon after uploading the video. We add this feature into the TSVM model, and get the classification result with the linear kernel. Comparing the false negative rate with the previous one, this classifier has better performance.


\subsubsection{Detecting IP generating fake views}
By using the video entropy analysis, we can improve anomaly detection for the IP dimension. As we discussed previously, detecting a fake view user behind a NAT box is more difficult. To detect fake view (NAT) IPs, we check whether the videos that IP viewed contain fake view videos, based on the fake view video detection methods in the last subsection. If a NAT IP has a great number of views, and many of them are watching the videos with low entropy values on the video dimension, We now have an effective method to classify the NAT IPs that have some fake view user \emph{hiding} behind it.

\begin{figure}[ht]
\centering
\includegraphics[width=0.6\textwidth]{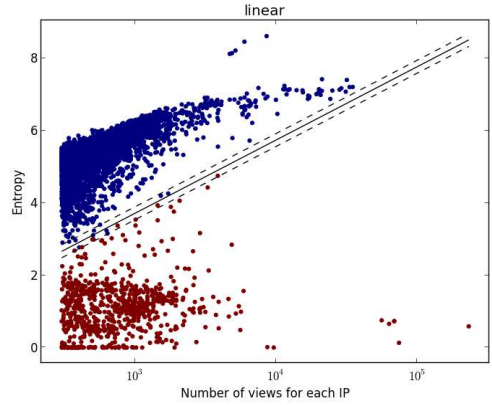}
\caption{TSVM classification for fake view detection on IP dimension.}\label{fig:svm_ip}
\end{figure}

We can also use the machine learning approach to improve the classification of fake view IPs. For IPs whose entropy values are near 0, we label them as fake view IPs. For IPs who have large entropy values, we label them as normal IPs. For IPs with middle entropy values and high view count, we will check if sufficient views are for fake view video(s). We classify a NAT IP with fake views as a fake view IP. As shown in Fig.~\ref{fig:svm_ip}, the fake view IPs can be distinguished with the hyperplane derived by the TSVM model.

\begin{figure}[ht]
\centering
\includegraphics[width=0.6\textwidth]{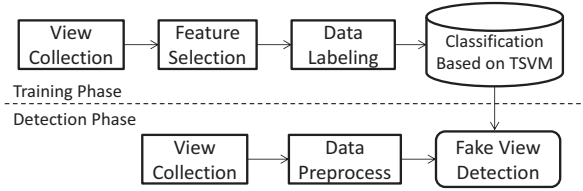}
\caption{Fake view detection.}\label{fig:process}
\end{figure}

We implement the fake view detection in two phases as summarized in Fig.~\ref{fig:process}. 1) We collect daily view records, and label a portion of the data based on the effective features as we discussed previously. By means of the semi-supervised learning approach, we get a TSVM model to classify the fake view videos and IPs in sequence. 2) In the detection phase, we use the classifier to get fake views out of all views after preprocessing the collected data. Also, we are able to evaluate the detection accuracy and error rate with labeled test dataset in this phase.

Based on the ``ground truth'' we gathered from the data with user IDs, we check if their IP addresses are the fake view IPs detected by the TSVM classifier. For performance evaluation, we observe that \textbf{99.08\%} of fake view IPs with users IDs are discovered in the fake view IP detection process. The TSVM classification performs well.

Finally, we give a more qualitative explanation why the video entropy is complementary to IP entropy. For IP entropy, if a particular IP $i$ has a single user repeatedly requesting for some particular video $k$, the IP entropy will be reduced quickly, since $(a_{i1}/w_i,\dots, a_{in}/w_i)$ becomes very skewed quickly due to the increase of $a_{ik}$. From the video entropy point view, however, an increased number of requests from an IP $i$ may not affect video entropy as much, since the skewness of $(a_{1k}/v_k,\dots, a_{ik}/v_k,\dots, a_{nk}/v_k)$ is unlikely to be affected significantly by $a_{ik}/v_k$ if $a_{ik}$ is generated by a single user. From the view point of a fake view user, it is less noticeable to generate fake views through some NAT IP $i$, since $w_i$ is large such that the abnormal increase of some $a_{ik}$ would not significantly reduce the IP entropy. Nonetheless, from the video entropy point of view, the $a_{ik}$ is a significant value compared with number of views from other IPs, i.e. $a_{hk}$, $h\neq i$. That means the sharp increase of $a_{ik}$ will reduce the video entropy $H_v(k)$. In summary, IP entropy together with video entropy can give a strong condition for detecting fake views.

\section{Online Detection of Fake Views}
Online detection of fake views is high desirable, but very challenging. It is well-known that YouTube needs to \emph{pause} the view count when it reaches 300 views, so that potential anomaly\footnote{In this case, the anomaly YouTube seems to check for is misleading titles for the video, which is a much harder problem which probably cannot avoid manual checking.} can be checked offline. Such offline checking of fake views caused much discussion among the service provider and users.

The difficulty with online checking and detection of fake views is mostly due to the scale of the video systems at hand. For the system deployed by our collaboration, it serves thousands of video requests per second at the peak load. At this scale, one needs a distributed data stream processing system\footnote{Such as the Twitter Storm and S4.} to handle the realtime processing requirements. We feel the simple entropy approach can be adapted and implemented in this case, and provide a rough online detection solution.

If even the entropy approach is too heavy-duty, a simpler online algorithm can also be adopted to catch only flagrant fake view abusers. In this case, it is possible to keep some hash tables completely in memory, keeping track of only the \emph{number of views} and the \emph{number of videos} an IP accesses. Based on these simple attributes, a rule-based approach can be applied to catch serious offenders. A simple rule may be expressed as simple as follows:
\[
views > K \wedge videos < L \Rightarrow \text{fake view}
\]
Other suspicious offenders can be flagged for more offline analysis. This kind of simple mechanisms makes a tradeoff between the detection effectiveness and the complexity of realization.

\section{Implementation and Case Study}
Assisted by our collaborator, we implemented our design for the view anomaly detection in their real-world VoD system. The fake view intruders are discovered based on our approach. By checking their behavior in data reports, we can validate the effectiveness of our fake view detection. The real experiments make our approach more credible for application in practice.

\begin{figure}[ht]
\centering
\includegraphics[width=0.6\textwidth]{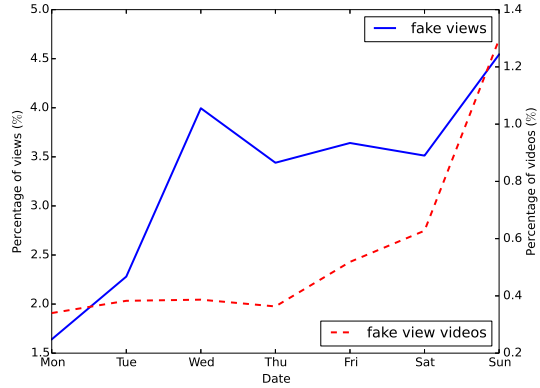}
\caption{Fake view detection result in one week.}\label{fig:case}
\end{figure}

We summarize some statistics of the detected fake views based on one week's data. Around 5\% videos get the number of views greater than 100 in each day, and their total views account for 95\% of the daily views. We select the videos with more than 100 views, and implement the fake view detection algorithm. As shown in Fig.~\ref{fig:case}, the fake views account for around 2\% $\sim$ 5\% of daily views (for videos with more than 100 views). And the fake view videos are usually less than 1\% of these videos viewed per day. Although this is not a high amount, as we explained earlier, it can at least give the VoD service provider a more accurate business/operational analysis by stripping the fake views from their data.

We observe that most fake view videos are UGC and MV (account for more than 90\% fake view videos), but also some popular TV series (usually the first episode). For UGC, it is because many video creators have great incentive to promote their uploaded videos. For MV, it may involve public relation companies and fans to introduce the fake views. We notice that the anomalous daily view pattern used as an example (in Section 2.4) is actually contributed by MV videos of one singer. Its attack was from multiple IPs. We conjecture that the intruder used a distributed network to inject the fake views. We show some features of representative samples of fake view videos in Table \ref{tab:vid}. Video1 got 10552 views from one single IP. Video2 was visited by 162 IPs, but 99.95\% views are out of 6 IPs. For video3, 10 IPs contributed 63.5\% views.
\begin{table}[htb]
\centering
\caption{Case study for fake view videos.}\label{tab:vid}
\begin{tabular}{c|c|c|c|c}
\hline
& Type & \# of Views & \# of IP & Entropy\tabularnewline
\hline
\hline
video1 & UGC & 10552 & 1   & 0 \tabularnewline
\hline
video2 & MV & 409409 & 162 & 1.62\tabularnewline
\hline
video3 & TV & 1388461 & 219584 & 5.02 \tabularnewline
\hline
\end{tabular}
\end{table}

Finally, we show some examples of fake view intruder detection in the IP dimension. IP1 viewed a lot, spread over 3 videos. IP2 had 73347 views, 97\% of which are on the same video. IP3 viewed 6366 videos, and ten of them account for 30\% views from this IP. These characteristics are typical of the fake view intruders. We validate the effectiveness of detection approach on one month dataset (around 1.2TB in size) assisted by a distributed data processing system.
\begin{table}[htb]
\centering
\caption{View anomaly detection example for IPs.}\label{tab:ip}
\begin{tabular}{c|c|c|c|c}
\hline
& Type & \# of Views & \# of Videos & Entropy\tabularnewline
\hline
\hline
IP1 & single & 228701  & 3   & 0.59  \tabularnewline
\hline
IP2 & single & 73347   & 4   & 0.13 \tabularnewline
\hline
IP3 & NAT    & 6366   & 1415 & 5.71 \tabularnewline
\hline
\end{tabular}
\end{table}

\section{Related Works}
Network anomaly detection attracted a lot interest from commercial concerns as well as academic research. Many works \cite{bp01imc,la04imc,mv03ac} study the flow level and traffic level anomalies for network monitoring and network security. In contrast, our work focus on the view count analytic issue at the application layer for online video service.

There appears to be similar issues whenever one wants to attract eyeballs on the Internet. \cite{cq12nsdi} detects fake accounts on online social networks (OSN). It ranks users according to their perceived likelihood of being fake based on social graph properties. By deploying their method on the largest OSN of Spain, $\sim$90\% of the 200K accounts are discovered as most likely to be fake.

The use of entropy function has been proposed for anomaly detection problems before. In \cite{lw01sp}, the authors propose to use several information-theoretic measures for network anomaly detection. The entropy measures are applied to Unix system call data, BSM data, and network tcpdump data to illustrate the utilities. \cite{gy05imc} uses two-phase entropy measures to detect network anomalies by comparing the current network traffic against a baseline distribution. A Maximum Entropy principle is applied to estimate the distribution for normal network operation using pre-labeled training data. Then the relative entropy of the network traffic is applied with respect to the distribution of the normal behavior. \cite{la06sigm} proposes efficient streaming algorithms to implement the entropy measurement on high-speed links with low CPU and memory requirements.

Machine learning approaches could be applied in many anomaly detection scenarios. Chandola et al. provide a survey of anomaly detection in \cite{cv09acs}. \cite{hp05sigc} uses the supervised learning to realize a mapping of traffic to applications based on labeled measurements from known applications. \cite{er08csa} proposes an improved K-means approach to classify unlabeled data into different categories for the anomaly intrusion detection. \cite{sr10sp} identifies the challenges for the intrusion detection community to employ machine learning effectively, and provides a set of guidelines for improvement.

Statistical approaches and signal analysis approaches are proposed for network anomaly detection. \cite{br02ss} presents a review of statistical fraud detection. \cite{wh02info} detects SYN flooding attacks based on the dynamics of the differences between the number of SYN and FIN packets, which is modeled as a stationary ergodic random process. In \cite{bp02imc}, Barford et al. apply wavelet techniques to aggregated traffic data in network flows for network traffic anomaly detection. In \cite{yd07smc}, Yeung et al. develop a covariance matrix method to model and detect flooding attacks. In \cite{sa05imc}, Soule et al. develop a traffic anomaly detection scheme based on Kalman Filter. \cite{bp02imc} reports results of signal analysis of four classes of network traffic anomalies: outages, flash crowds, attacks and measurement failures. In \cite{tm03sp}, authors propose to use the wide array of signal processing methods to solve the problem of anomaly detection. It seems more collaboration between the networking and signal processing will help develop better and more effective tools for detecting network anomalies and performance problems.

\section{Conclusion}

We are fortunate that our collaborator, Tencent Video, allows us to study this interesting problem using data from their online VoD system. Our study is focused on fake views caused by robots generating fake video requests or reports. Our study found effective methods based on the use of IP entropy and video entropy functions, together with other strong features such as publish date of a video. We found an interesting way to use IP entropy and video entropy together to make our technique more effective. We also report some fake view statistics found in a real-world online VoD system.

There are still many challenging problems for further study. Developing online algorithms is very desirable and still requires further work. Cracking the more challenging attacks involving multiple IPs (botnets) is also very challenging.



\section*{Acknowledgments}
We would like to thank Tencent Video staff for providing the generous data support, and assisting us to evaluate our proposal.

\bibliographystyle{abbrv}
\bibliography{main_bib}

\end{document}